**Title**

Self-Shielding Enhanced Organics Synthesis in an Early Reduced Earth's Atmosphere

**Authors**

Tatsuya Yoshida[1], Shungo Koyama[1], Yuki Nakamura[2], Naoki Terada[1], and Kiyoshi Kuramoto[3]

1: *Graduate School of Science, Tohoku University, Sendai, Miyagi 980-8578, Japan*

2: *Graduate School of Science, University of Tokyo, Bunkyo, Tokyo 113-0033, Japan*

3: *Faculty of Science, Hokkaido University, Sapporo, Hokkaido 060-0810, Japan*

**Abstract**

Earth is expected to have acquired a reduced proto-atmosphere enriched in $H_2$ and $CH_4$ through the accretion of building blocks that contain metallic Fe and/or the gravitational trapping of surrounding nebula gas. Such an early, wet, reduced atmosphere that covers a proto-ocean would then ultimately evolve toward oxidized chemical compositions through photochemical processes that involve reactions with $H_2O$-derived oxidant radicals and the selective escape of hydrogen to space. During this time, atmospheric $CH_4$ could be photochemically reprocessed to generate not only C-bearing oxides but also organics. However, the branching ratio between organic matter formation and oxidation remains unknown despite its significance on the abiotic chemical evolution of early Earth. Here, we show via numerical analyses that UV absorptions by gaseous hydrocarbons such as $C_2H_2$ and $C_3H_4$ significantly suppress $H_2O$ photolysis subsequent $CH_4$ oxidation during the photochemical evolution of a wet proto-atmosphere enriched in $H_2$ and $CH_4$. As a result, nearly half of the initial $CH_4$ converted to heavier organics along with the deposition of prebiotically essential molecules such as HCN and $H_2CO$ on the surface of a primordial ocean for a geological timescale order of 10-100 Myr. Our results suggest that the accumulation of organics and prebiotically important molecules in the proto-ocean could produce a soup enriched in various organics, which might have eventually led to the emergence of living organisms.

## 1. Introduction

Recent cosmochemical studies that have compared isotopic compositions among Earth's materials and primitive meteorites indicate that most of Earth's accreting materials closely resembled enstatite chondrites, which have the most reduced oxidation state among primitive meteorites, throughout the accretion phase (e.g., Dauphas, 2017; Dauphas et al., 2024; Halliday and Canup, 2023 and references therein). This suggests that the volatile elements accreted on Earth were chemically reduced by reactions with metallic Fe in the accreting materials, forming an impact-generated atmosphere enriched in reduced gases such as $H_2$ and $CH_4$ (Urey, 1952; Kuramoto and Matsui, 1996; Schaefer and Fegley, 2010; Genda et al., 2017a; Genda et al., 2017b; Benner et al., 2020; Zahnle et al., 2020; Itcovitz et al., 2022; Pearce et al., 2022; Wogan et al., 2023). In addition, planet formation theories and the primordial isotopic compositions of hydrogen and noble gases in igneous rocks derived from the deep mantle suggest that proto-Earth also obtained the $H_2$-dominated solar nebula gas as its proto-atmosphere (Hallis et al., 2015; Olson and Sharp, 2019; Saito and Kuramoto, 2020).

Such a highly reduced atmosphere covering a proto-ocean has been thought to be a candidate site for the synthesis of organic matter and prebiotically important molecules such as HCN and $H_2CO$ that precede the emergence of life (e.g., Miller, 1953; Miller and Urey, 1959; Schlesinger and Miller, 1983; Benner et al., 2020; Zahnle et al., 2020; Pearce et al., 2022; Wogan et al., 2023). These molecules can be produced through chemical reactions induced by the irradiation of high-energy photons and particles (e.g., Zahnle, 1986; Airapetian et al., 2016), electric discharge (e.g., Miller, 1953; Schlesinger and Miller, 1983), and/or high-velocity impacts (Furukawa et al., 2009; Ferus et al., 2020; 2022). Among these energy sources, solar UV irradiation is expected to be the most potentially available for the synthesis of these compounds (Chyba and Sagan, 1992). Synthesized HCN and $H_2CO$ could be precursors of amino acids and part of nucleobases and sugars: various amino acids can be produced from HCN, aldehydes, and $NH_3$ through Strecker reactions (Miller and Urey, 1959). Adenine, which is one of the nucleobases can be generated in the solution of HCN and $NH_3$ (Oró, 1960), and ribose, a building block of nucleotides, can be produced from $H_2CO$ through the formose reactions (Butlerow, 1861). $NH_2CHO$, produced through the hydrolysis of HCN, can serve as a precursor to all canonical nucleobases (e.g., Saladino et al., 2012). Organic haze aerosols produced in reduced atmospheres might be deposited on the surface and form a water-insoluble organics layer, part of which could hydrolyze to produce various amino acids and nucleobases (e.g., Khare et al., 1986; Poch et al., 2012; Pearce et al., 2024).

Atmospheric photochemistry that involves $H_2O$-derived oxidant radicals, with hydrogen escaping into space, may have evolved an early Earth's reduced atmosphere toward oxidized chemical compositions, although its rate was possibly slow. The hydrogen escape rate is estimated to asymptote to the diffusion-limited flux even under the enhanced X-ray and extreme ultraviolet (XUV) irradiation from the young Sun when considering the radiative cooling effects by $CH_4$ and radiatively active photochemically derived products (Yoshida and Kuramoto, 2020; 2021). This prolongs the duration of the $H_2$-rich condition, possibly up to several hundred million years, which is significantly longer than in pure hydrogen atmospheres (e.g., Sekiya et al., 1980; Erkaev et al., 2013; Lammer et al., 2014). On the other hand, $CH_4$ would become oxidized to CO and $CO_2$ through reactions that involve oxidant radicals in parallel with its photochemical conversion to HCN, $H_2CO$, and other heavier organics.

The production rate of oxidant radicals primarily supplied by $H_2O$ photolysis depends not only on $H_2O$ concentration but also on the availability of UV photons at $H_2O$ photolysis wavelengths. While UV shielding by organic haze may suppress molecular photolysis as suggested for the protection of $NH_3$ in ancient Earth's atmosphere (Sagan and Chyba, 1997), it remains unknown how low molecular weight organic compounds synthesized prior to the organic haze formation may affect the rates of $H_2O$ and $CH_4$ photolysis and subsequent chemical reactions. Specifically, UV shielding effects of gaseous hydrocarbons that contain more than 2 carbons have been given little consideration.

In the present study, we employed a 1-D atmospheric photochemical model for an early reduced Earth's atmosphere mainly composed of $H_2$ and $CH_4$ to clarify the UV shielding effects of gaseous hydrocarbons on the production of carbon oxides and organic matter, and photochemical evolution of the early Earth's atmosphere.

This contribution is organized as follows. In Section 2, we describe the outline of our photochemical model. In Section 3, we show the numerical results of the atmospheric profile, production/loss rates of each chemical species, and deposition rates of prebiotically important molecules. In Section 4.1, we show the comparison with the results without the UV absorption by hydrocarbons to clarify their UV shielding effects. In Sections 4.2 and 4.3, we analyze the effects of hydrocarbon accumulation and $CO_2$ degassing on the production and loss rates of major species. In Section 4.4, we discuss the dependence of the calculation results on uncertain parameters. In Section 4.5, we show possible evolutionary tracks of the early reduced Earth's atmosphere based on our calculation results.

## 2. Model description

We developed a 1-D photochemical model for a reduced Earth's atmosphere mainly composed of $H_2$ and $CH_4$ based on PROTEUS (Nakamura et al., 2023a). So far, PROTEUS has been successfully applied to the jovian ionosphere (Nakamura et al., 2022), the present-day martian atmosphere (Nakamura et al., 2023b; Yoshida et al., 2023), an early martian atmosphere (Koyama et al., 2024), and an $H_2O$-dominated atmosphere in the runaway greenhouse condition (Kawamura et al., 2024). The details of PROTEUS are described in the work of Nakamura et al. (2023a). Below, we present the outline of the model in this study.

As for the chemical processes, we consider 342 chemical reactions (Table S1) for 61 chemical species composed of H, O, C, and N; $H_2$, $CH_4$, $N_2$, $H_2O$, $O_2$, O, O($^1$D), H, OH, $O_3$, $H_2O_2$, $CO_2$, CO, $H_2CO$, HCO, $HO_2$, $^1CH_2$, $^3CH_2$, $CH_3$, $C_2H_6$, $C_2H_2$, $C_2H_4$, $HNO_2$, NO, $HNO_3$, $NO_2$, N, HNO, $NH_3$, $NH_2$, $N_2H_4$, $N_2H_3$, NH, $C_2H$, $C_2$, $C_3H_8$, $C_3H_6$, $C_2H_5$, $CH_2CCH_2$, CH, C, $CH_2CO$, $CH_3CHO$, $C_2H_5CHO$, $C_3H_3$, $C_3H_2$, $CH_3C_2H$, HCN, CN, HNCO, NCO, $C_3H_7$, $C_2H_3$, $C_3H_5$, $CH_3O_2$, $CH_3CO$, $C_2H_2OH$, $C_2H_4OH$, $CH_3O$, HCNOH, and $H_2CN$. Referring to the chemical reaction list of Tian et al. (2011), we consider C-bearing species with up to 3 carbons, excluding S-bearing species and organic haze aerosols, to clarify the role of UV shielding by gaseous hydrocarbons.

The model considers the vertical transport of each chemical species due to molecular diffusion and eddy diffusion. For the molecular diffusion coefficient, we use the formula derived by Banks and Kockarts (1973):

$$D_i = 1.52 \times 10^{18} \left(\frac{1}{M_i} + \frac{1}{M}\right) \frac{T^{0.5}}{n} \quad (1)$$

where $D_i$ is the molecular diffusion coefficient of species $i$ (cm$^2$ s$^{-1}$), $M_i$ is the molecular mass of species $i$, $M$ is the mean molecular mass of the atmosphere, $T$ is the temperature (K), and $n$ is the total number density of the atmosphere (cm$^{-3}$). The profile of the eddy diffusion coefficient, which depends on atmospheric pressure, is taken from the settings for a reducing atmosphere by Hu et al. (2012) (Figure 1). We discuss the effect of the change in the eddy diffusion coefficient on the calculated results in Section 4.4 since the eddy diffusion coefficient in the early atmosphere is highly uncertain.

To calculate the profiles of the photolysis rates, we adopt the UV spectrum from 100 to 1000 nm estimated for the young Sun at the age of 100 Myr (Claire et al., 2012), and solve the radiative transfer by considering the absorption of the solar irradiation by chemical species assuming the solar zenith angle to be 57.3 degrees as indicated by Hu et al. (2012). For the absorption cross-sections, we mainly refer to the MPI-Mainz-UV-VIS Spectral Atlas of Gaseous Molecules (Keller-Rudek et al., 2013) and the JPL publication (Burkholder et al., 2015).

The temperature profile at the troposphere is assumed to follow adiabat, and that above the tropopause is assumed to be kept at the skin temperature (Zahnle et al., 2020). Here, we put the tropopause at 0.1 bar as it is in most solar system planets with atmospheres (Robinson and Catling, 2014). The water vapor profile at the troposphere is assumed to follow the saturation vapor pressure as a function of temperature. As the standard case, the mixing ratio of water vapor at the stratosphere is fixed at 1 ppm by adjusting the planetary albedo and the skin temperature at 175 K, which is consistent with the stratospheric temperature under the faint young Sun (Tian et al., 2011; Zahnle et al., 2020). The mixing ratio of water vapor in the stratosphere is an important parameter in the photochemistry of the reduced atmosphere since water vapor is often the major oxidant. We discuss the effect of its change on the calculated atmospheric profile in Section 4.4.

At the lower boundary, which corresponds to the surface, the number densities of $H_2$, $CH_4$, and $N_2$ are fixed. Here, the basal number densities of $CH_4$ and $N_2$ are set to satisfy the column abundances equivalent to the amount of carbon on the present-day Earth's surface layer ($1.5 \times 10^7$ mol/m$^2$; Holland, 1984) and that of nitrogen molecules in the present-day Earth's atmosphere ($3.0 \times 10^5$ mol/m$^2$). The basal $CH_4/H_2$ ratio is taken to be a parameter. HCN, $H_2CO$, and $NH_3$ are assumed to be deposited on the surface at the constant deposition velocities of $7.0 \times 10^{-5}$, $1.0 \times 10^{-3}$, and $1.0 \times 10^{-2}$ m/s, respectively (Tian et al., 2011; Hu et al., 2012). The upper boundary is set at the altitude where the pressure becomes lower than 0.01 Pa and optically thin in the UV wavelength range considered by this study. At the upper boundary, atomic N is assumed to flow down at the constant downward flux supposing its production through the photolysis of $N_2$ in the thermosphere/ionosphere (Zahnle, 1986; Tian et al., 2011). The downward flux of N is given by

$$F_N = \frac{1}{2} \int_\lambda F_{UV}(\lambda) \frac{\sigma_{N_2}(\lambda) f_{N_2}/M_{N_2}}{\sum_i \sigma_i(\lambda) f_i/M_i} d\lambda \quad (2)$$

where $F_{UV}(\lambda)$ is the UV flux per wavelength, $\sigma_i(\lambda)$ and $f_i$ are the absorption cross-section and the basal volume mixing ratio of major UV absorbers $i$ (=$N_2$, $H_2$, $CH_4$, $C_2H_6$, CO, and $CO_2$). Here, the UV wavelength range absorbed by $N_2$ is from 0.5 nm to 103 nm. The dependence of the downward flux of N on the basal $CH_4/H_2$ ratio is shown in Figure 2. The N flux is about 4-5 orders of magnitudes lower than that supposed by Zahnle (1986) and Tian et al. (2011) because the photodissociation of $N_2$ is suppressed by surrounding $H_2$ and $CH_4$. Under these boundary conditions, the continuity-transport equations that involve chemical processes are solved by numerical integration over time. As the initial condition, the number densities of $H_2$, $CH_4$, and $N_2$ are set to be in hydrostatic equilibrium with a uniform composition. The integration is done over a period at least longer than the eddy diffusion timescale (~$10^9$ s), and the abundance of the C-bearing chemical products from $CH_4$ reaches 0.1% of that of $CH_4$. The integration is completed when the distribution of short-lived species settles into a steady state and the production/loss rates of the long-lived species become nearly constant. The cases where long-lived chemical products significantly accumulate in the atmosphere are described in Section 4.2.

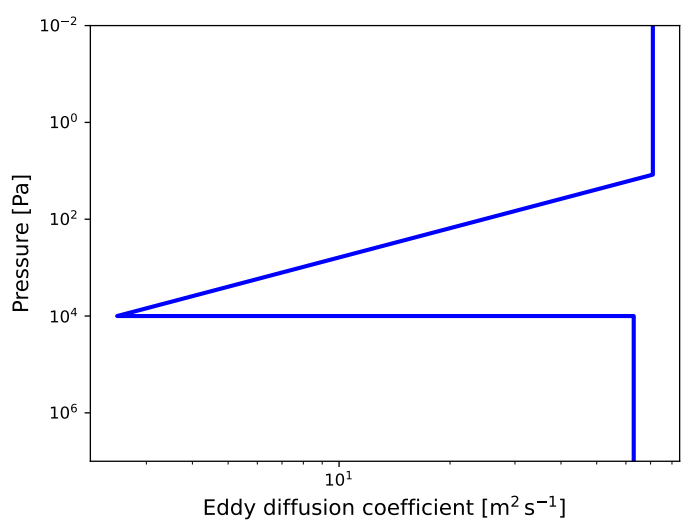

Figure 1. Eddy diffusion coefficient profile.

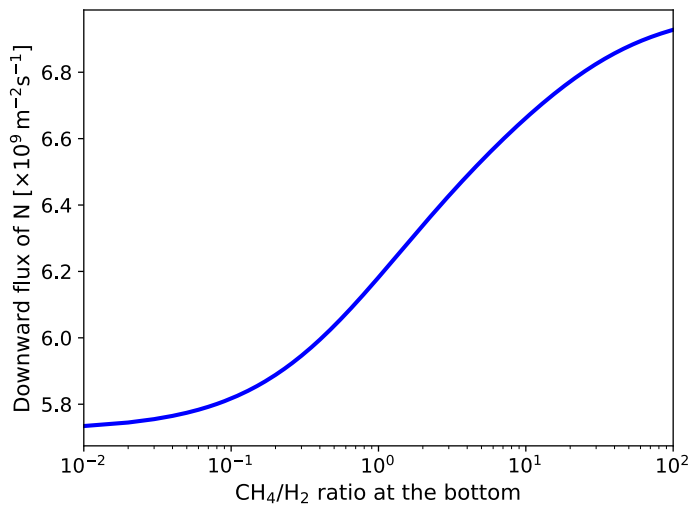

Figure 2. Downward flux of N depending on the basal $CH_4/H_2$ ratio.

## 3. Results

The number density profiles are shown in Figure 3 for basal $CH_4/H_2$ ratios of 0.1 and 10. Here, various C-bearing species, including both organics and oxides, are produced from $CH_4$. While the abundance of hydrocarbons such as $C_2H_6$ and $C_3H_8$ is large when $CH_4/H_2=10$, C-bearing oxides such as CO and $CO_2$ are depleted in this case compared with those in the $H_2$-dominated condition.

As indicated by the number density profiles, the production rates of hydrocarbons increase and far exceed those of carbon oxides under higher $CH_4/H_2$ ratios (Figure 4(a)). As the mixing ratio of $CH_4$ increases, the production of hydrocarbons increases with the net $CH_4$ decomposition rate (Figure 4(a)(b)). Concurrently, the production of oxidant radicals by photolysis of $H_2O$ vapor, given saturated at the tropopause level, becomes suppressed (Figure 5(a)) because the photochemically produced hydrocarbons such as $C_2H_2$ and $CH_2CCH_2$ effectively shield UV in wavelength range photolyzing $H_2O$ as shown in the absorption cross-section of each species in Figure 5(b).

Atmospheric escape would progressively increase the $CH_4/H_2$ ratio in the proto-atmosphere because $CH_4$ escape is likely negligible (Yoshida and Kuramoto, 2021), and the $H_2$ loss would proceed with a timescale order of 10-100 Myr, which is shorter than the timescale of photochemical $CH_4$ loss. $CH_4$ is more stable, particularly in early $H_2$-rich conditions: the timescale for $CH_4$ loss obtained from the $CH_4$ column density divided by its net column decomposition rate is as long as ~1 Gyr when $CH_4/H_2$~0.01. This long timescale primarily results from the effective $CH_4$ reformation reactions, such as $CH_3+H+M \rightarrow CH_4+M$ (R110) (Figure 4(b)), which is well-known for the mesosphere regions of giant planets (Yung and DeMore, 1998). The effectiveness of the $CH_4$ reformation reactions on $H_2$-rich atmospheres, supposing primitive Earth, is also shown by Wogan et al. (2023). Moreover, the reactions that oxidize $CH_4$ such as $OH+CH_4 \rightarrow CH_3+H_2O$ (R96) are suppressed in $H_2$-rich conditions (Figure 4(b)) because OH is depleted by the reaction with $H_2$ (R62: $H_2+OH \rightarrow H_2O+H$), which also contributes to stabilizing $CH_4$. As the atmospheric $CH_4/H_2$ ratio increases through the hydrogen escape to space, the photochemical production of organics by the polymerization of $CH_4$-derived species becomes enhanced due to the slowing of hydrogenation reactions.

$H_2CO$ and HCN, which are considered crucial in prebiotic chemical evolution, can be continuously synthesized and accumulated in the proto-ocean during the evolution of the reduced atmosphere. In our photochemical model that contains background $N_2$ equivalent to 0.8 bar at the surface, the deposition rate of $H_2CO$ peaks at around $CH_4/H_2=1$ (Figure 4(c)), where hydrogenation of HCO proceeds efficiently near the surface. The maximum deposition rate of $H_2CO$ is at most ~$10^{14}$ molecules m$^{-2}$ s$^{-1}$, slightly lower than estimates for a weakly reduced $N_2$-dominated atmosphere ($2.8 \times 10^{14}$ molecules m$^{-2}$ s$^{-1}$; Pinto et al., 1980) mainly due to the depletion of CO, which is a source of HCO. When CH$_4$/H$_2$>~0.1, HCN becomes the main compound formed directly by reactions that involve atomic N due to the abundant CH$_3$ in the upper atmosphere, whereas NH$_3$ production is kept slower in the whole range of the considered atmospheric composition. The deposition rate of HCN reaches ~10$^{10}$ m$^{-2}$ s$^{-1}$ under high CH$_4$/H$_2$ conditions (Figure 4(c)) but is significantly lower than the maximum value of ~10$^{14}$ m$^{-2}$ s$^{-1}$ in an N$_2$-dominated atmosphere (Zahnle, 1986; Tian et al., 2011). This low deposition rate is mainly because of the suppressed downward flux of atomic N from the upper boundary due to the inhibition of N$_2$ photolysis by UV absorption by abundant H$_2$, CH$_4$, and other hydrocarbons (Figure 2). Although the deposition rates of H$_2$CO and HCN tend to be lower than those in an N$_2$-dominated atmosphere, their long-term deposition could result in their concentrations sufficient to produce amino acids and nucleobases in the proto-ocean as described in Section 4.5.

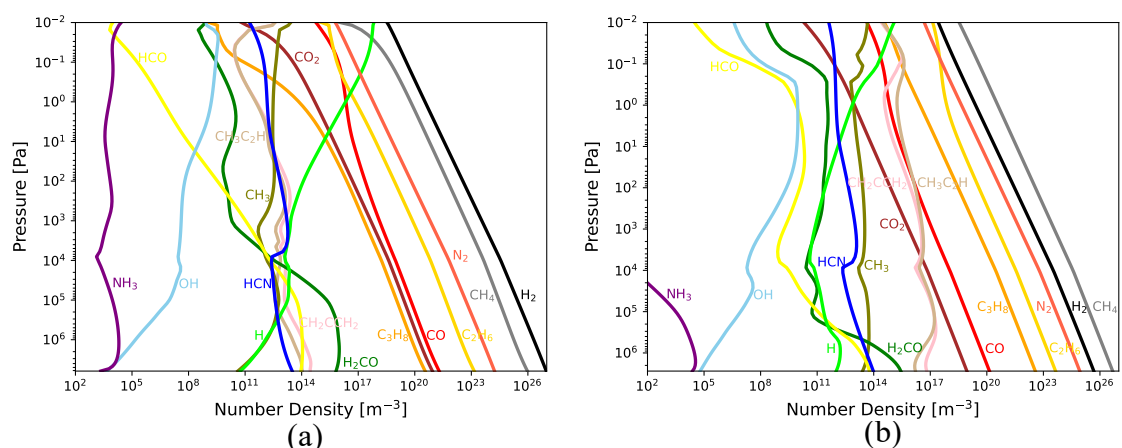

Figure 3. Number density profiles when the basal CH$_4$/H$_2$ ratio is 0.1 (a) and 10 (b).

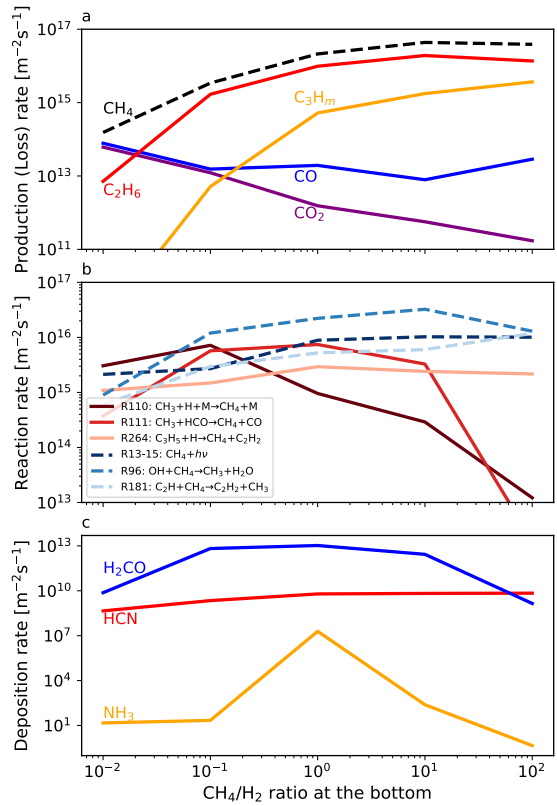

Figure 4. (a) Column-integrated production/loss rates of $CH_4$, hydrocarbons, and oxides depending on the basal $CH_4/H_2$ ratio. The black dashed line represents the net loss rate of $CH_4$. The solid lines represent the net production rate of each chemical species. "$C_3H_m$" represents the hydrocarbons with 3 carbons. (b) Column-integrated rates of the chemical reactions related to the production (red solid lines) and loss (blue dashed lines) of $CH_4$. (c) Deposition rates of $H_2CO$, HCN, and $NH_3$.

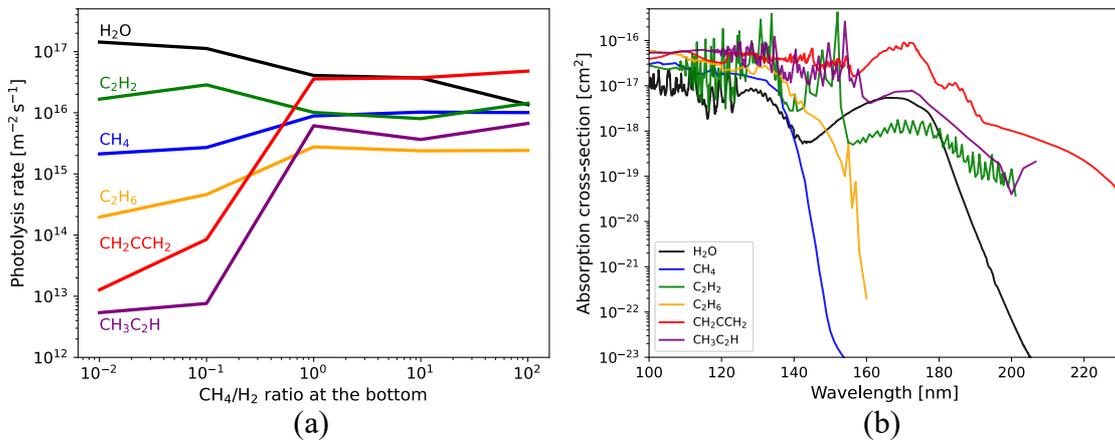

Figure 5. (a) Column-integrated photolysis rate of major UV absorbers. (b) Absorption cross-sections with UV wavelength.

## 4. Discussion

### 4.1. Comparison with the results without the UV absorption by hydrocarbons

To clarify the UV shielding effects of hydrocarbons, the comparison with the simulation results without the UV absorption by hydrocarbons other than $CH_4$ (Figure 6(a)) and $C_3H_m$ neglected by previous studies such as Pearce et al. (2022) and Wogan et al. (2023) (Figure 6(b)) are shown. The production rate of carbon oxides increases while that of $C_3H_m$ relatively decreases without UV absorption by hydrocarbons since the formation of oxidant radicals by $H_2O$ photolysis becomes enhanced. The UV shielding by $C_3H_m$ becomes effective, especially in $CH_4$-dominated conditions (Figure 6(b)). The production of $C_2H_6$ is a little enhanced in Figure 6(a) due to the more efficient $CH_4$ photolysis. These results confirm that UV shielding by gaseous hydrocarbons significantly increases the branching ratio of organics production through $CH_4$ photolysis compared with estimates by the previous studies.

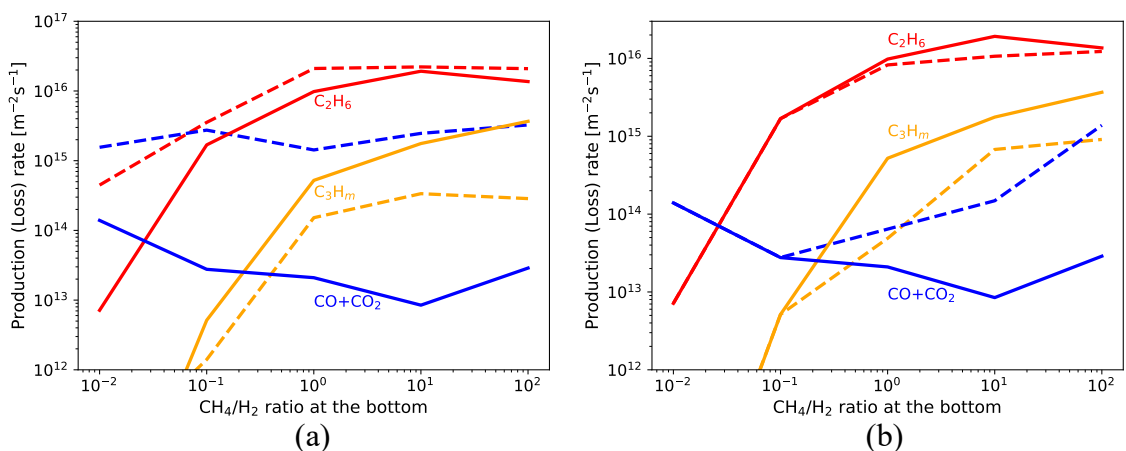

Figure 6. (a) Column-integrated production rates of each major species as a function of the basal $CH_4/H_2$ ratio. The solid lines represent the production rates with UV absorption by hydrocarbons included, while the dashed lines represent the production rates without UV absorption by hydrocarbons other than $CH_4$. (b) Same as (a) but the results without UV absorption by hydrocarbons with 3 carbons ($C_3H_m$). The solid lines are the same as those on (a) and the dashed lines are the results without the UV absorption by $C_3H_m$.

## 4.2. Effects of hydrocarbons accumulation

Our results show that hydrocarbons are efficiently produced from $CH_4$ due to their UV shielding effects. The production rate of $C_2H_6$ is especially high among C-bearing chemical products (Figure 4(a)). Since $C_2H_6$ is not a radical species, it may accumulate in the proto-atmosphere. The effects of the accumulation of $C_2H_6$ on the production and deposition rates of other relevant chemical species are examined through simulations with various atmospheric $C_2H_6$ abundances, assuming a fraction of $CH_4$ is converted to $C_2H_6$ while fixing the total carbon amount equivalent to the present-day Earth's surface layer. $C_2H_6$ is photolyzed to generate mainly heavier hydrocarbons such as $C_3H_8$ when the basal $C_2H_6/CH_4$ is larger than ~0.1 (Figure 7(a)). The production rates of CO and $CO_2$ are relatively small compared with that of heavier hydrocarbons due to the UV shielding by hydrocarbons as in $CH_4$-rich conditions. The deposition rates of HCN, $H_2CO$, and $NH_3$ are comparable with those in $CH_4$-dominated conditions (Figure 7(b)). These results indicate that effective organics production continued even after $C_2H_6$ accumulation proceeded as shown in the atmospheric evolutionary track in Section 4.5.

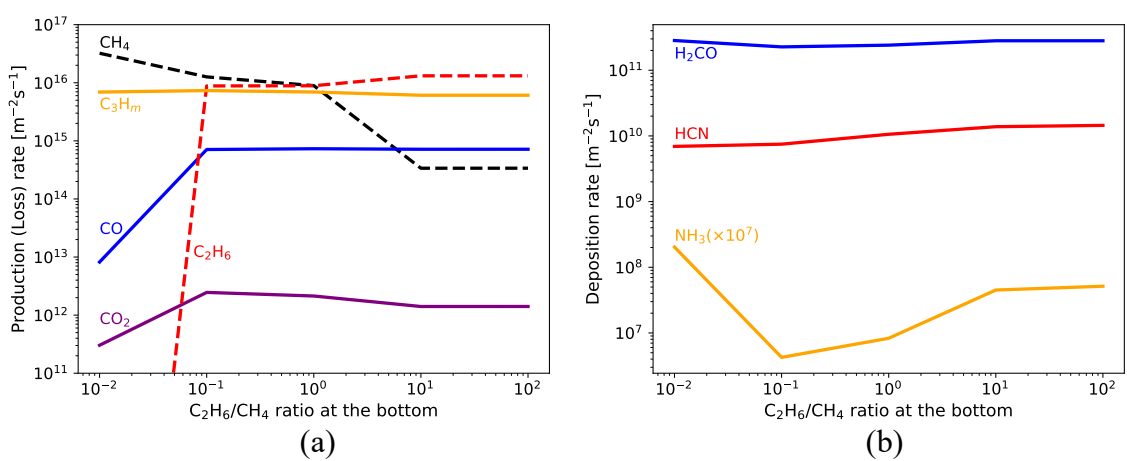

Figure 7. (a) Column-integrated production/loss rates of each chemical species depending on the basal $C_2H_6/CH_4$ ratio. The solid and dashed lines represent the net production and loss rates, respectively. "$C_3H_m$" represents the hydrocarbons with 3 carbons. (b) Deposition rates of HCN, $H_2CO$, and $NH_3$ depending on the basal $C_2H_6/CH_4$ ratio.

## 4.3. Effects of $CO_2$ supply

We suppose that the initial atmosphere is mainly composed of $H_2$ and $CH_4$. On the other hand, oxidized gases such as $CO_2$ may have been provided by volcanic degassing if the upper mantle had a similar redox state to today. The effects of $CO_2$ supply on the production/loss rates of major chemical species are shown by simulations given various basal $CO_2/CH_4$ ratios in Figure 8. The CO production rate increases with the $CO_2/CH_4$ ratio (Figure 8(a)). $CO_2$ photolysis becomes the major source of oxidant radicals when $CO_2/CH_4 > \sim 10^{-5}$, which can be maintained by $CO_2$ degassing with rates higher than the present-day value ($\sim 10^{13}$ m$^{-2}$ s$^{-1}$; Catling and Kasting, 2017). In these $CO_2$-rich conditions, about half of CO is derived directly from $CO_2$ photolysis, and the other half is produced through $CH_4$ oxidation by oxidant radicals mainly provided from $CO_2$ photolysis. CO becomes the major chemical product when $CO_2/CH_4 > \sim 0.01$, while the $CO_2$ degassing rate needs to be more than 2 orders of magnitude higher than the present-day value to maintain the $CO_2$ abundance. The result that CO becomes the major $CH_4$-derived chemical product in $CO_2$-dominated conditions is consistent with the photochemical calculation by Wogan et al. (2023) for an $H_2$-$CH_4$-$CO_2$ atmosphere. These results indicate that the $CH_4$ oxidation may have been promoted by volcanic $CO_2$ supply as long as the supply was high enough to make $CO_2$ the main source of oxidant radicals, although the degassing rate from proto-Earth is highly uncertain.

Efficient production of organics by their UV shielding occurs even in $CO_2$-rich conditions; their production rate little changes in the wide atmospheric composition range with $CO_2/CH_4 < \sim 1$ (Figure 8(a)). To clarify the effects of UV absorption by $C_3H_m$ neglected by previous studies, the comparison with the simulation results without their UV absorption is shown in Figure 8(b). The production rate of CO becomes high, while that of $C_3H_m$ is significantly suppressed compared with the standard case due to the enhancement of $H_2O$ photolysis, which shows that UV shielding by $C_3H_m$ can enhance organics production in $CO_2$-rich conditions. The production rate of $C_2H_6$ is slightly large without UV shielding by $C_3H_m$ because of the more efficient $CH_4$ photolysis.

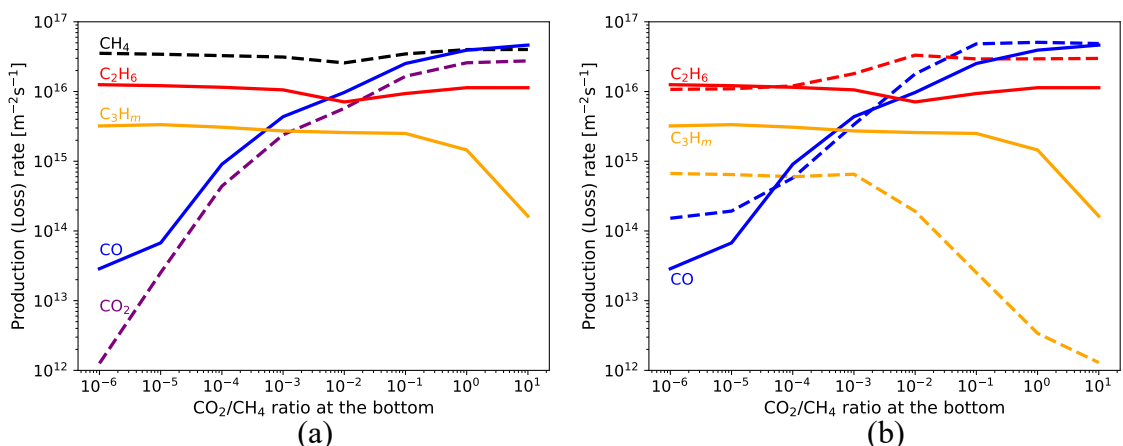

Figure 8. (a) Column-integrated production/loss rates of each major species depending on the basal $CO_2/CH_4$ ratio. Here, the basal $CH_4/H_2$ ratio is 10. The solid and dashed lines represent the net production and loss rates, respectively. (b) Same as (a) but the results without UV absorption by hydrocarbons with 3 carbons ($C_3H_m$). The solid lines are the same as those on (a) and the dashed lines are the results without the UV absorption by $C_3H_m$.

## 4.4. Parameter dependence

### 4.4.1. Dependence on the eddy diffusion coefficient

The profile of the eddy diffusion coefficient in the early Earth's upper atmosphere is highly uncertain, although the setting should affect the calculation results. Figure 9(a) shows the comparison with the production/loss rates of major species under the eddy diffusion coefficient 10 times as large as the standard setting. In the higher eddy diffusion condition, the abundance of $CH_4$ in the uppermost region increases due to the efficient vertical transport. This slightly enhances the $CH_4$ photolysis and the following production of hydrocarbons. The production of C-bearing oxides is also slightly enhanced when the basal $CH_4/H_2$ ratio is small but becomes suppressed as the $CH_4/H_2$ ratio increases. The overall dependence of the production of $CH_4$-derived major products on the atmospheric composition is little affected by the increased eddy diffusion, which indicates that the UV shielding effect of hydrocarbons plays a primary role in controlling this behavior.

**4.4.2. Dependence on the stratospheric $H_2O$ abundance**

Since $H_2O$ is the major source of oxidant radicals through its photolysis, its stratospheric abundance is a key parameter in the oxidation of reduced atmospheres. We suppose that the stratospheric $H_2O$ mixing ratio is 1 ppm, which is slightly lower than the present value, considering a colder stratosphere under the faint young Sun. On the other hand, the stratospheric temperature and $H_2O$ mixing ratio are variable depending on the uncertain planetary albedo on early Earth. Figure 9(b) shows the production/loss rates of major species when the stratospheric $H_2O$ mixing ratio is 10 times as large as the standard setting. The production rates of CO and $CO_2$ increase in the abundant stratospheric $H_2O$ condition due to the increase in oxidant radicals such as OH and O. Despite the production enhancement of C-bearing oxides, however, the efficient production of hydrocarbons has little changed in the $CH_4$-rich conditions due to their strong UV shielding effect.

**4.4.3. Dependence on the UV flux**

While our nominal simulation used a fixed UV flux estimated for the young Sun at the age of about 100 Myr, they should have changed with time. In particular, the flux of UV with a wavelength shorter than about 200 nm is estimated to decline rapidly (Claire et al., 2012). Figure 9(c) shows the production/loss rates of major chemical species when the whole UV flux is half as high as the standard case with the fixed spectrum profile. The net loss rate of $CH_4$ decreases proportionally with the UV flux, while the relative production rates of major chemical products little change. Considering the change in the UV flux with time, the timescale for $CH_4$ decomposition should be prolonged compared with the results in Section 3.

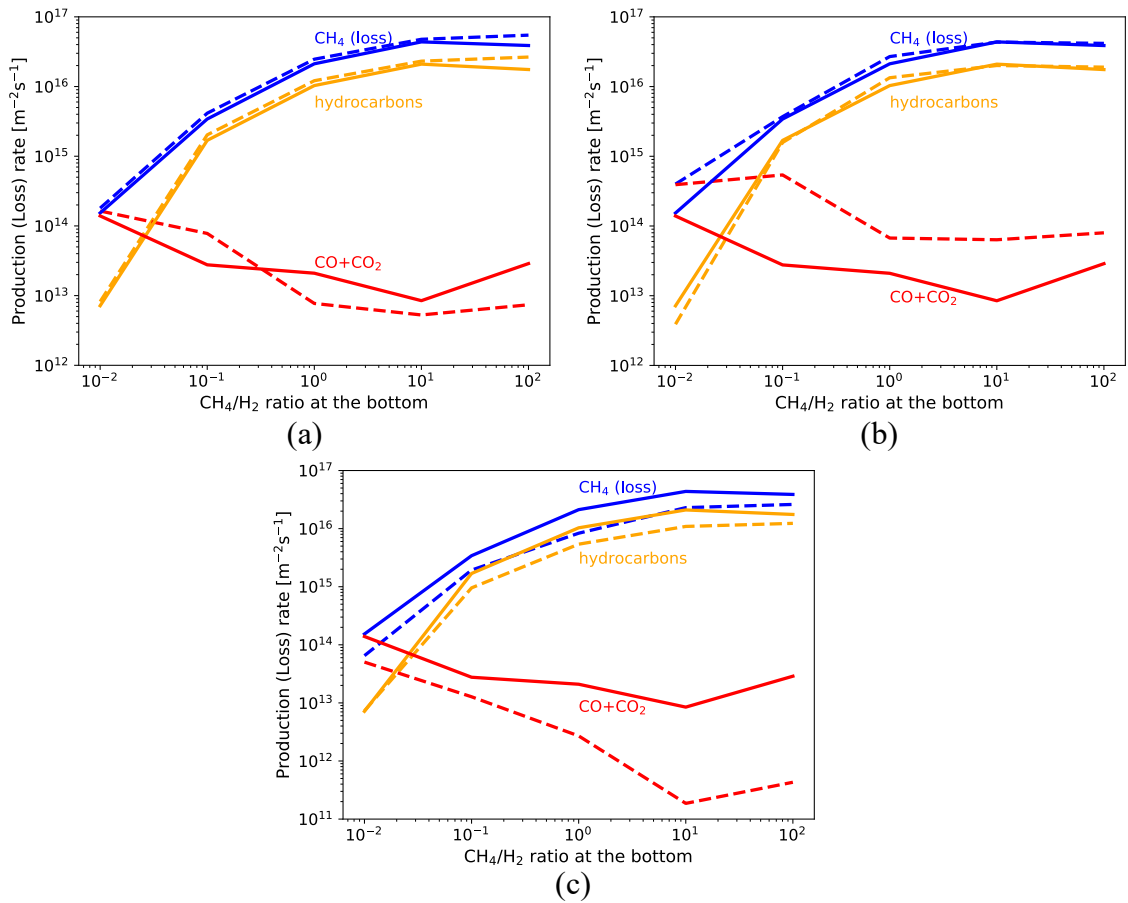

Figure 9. (a) Column-integrated production/loss rate when the magnitudes of the eddy diffusion coefficient are the standard setting (Solid lines) and 10 times as large as the standard setting (Dashed lines). (b) Column-integrated production/loss rates when the stratospheric $H_2O$ mixing ratios are 1 ppm (Solid lines) and 10 ppm (Dashed lines). (c) Column-integrated production/loss rates under the standard UV flux setting (Solid lines) and those under the UV flux half as large as the standard setting (Dashed lines).

## 4.5. Evolution of early reduced atmosphere on Earth

Based on the production and loss rates of chemical species calculated with varying background atmospheric composition, possible evolutionary tracks are drawn for the early reduced Earth's atmosphere. The initial atmosphere is composed of $H_2$ $CH_4$, and 0.8 bar $N_2$ with taking the amount of $CH_4$ equivalent to the C content in the present Earth's surface layer (=$1.5 \times 10^7$ mol/m$^2$, convertible to the $CH_4$ load of 20 bar on Earth's surface) including crustal reservoirs (Holland, 1984) and $CH_4/H_2$ ratio to be a parameter. $H_2$ is supposed to escape to space with diffusion-limited escape rates (Yoshida and Kuramoto, 2021). The stratospheric $H_2O$ mixing ratio is fixed at 1 ppm. The temporal change in the abundance of long-lived species $H_2$, $CH_4$, $C_2H_6$, CO, and $C_3H_m$ follows their production/loss rates calculated for each evolutionary step. The deposition of $H_2CO$, HCN, and $NH_3$ are simultaneously obtained assuming no return flux of them from the proto-ocean to the atmosphere. For simplicity, we neglect possible volatile delivery and other atmospheric escape processes during atmospheric evolution.

An example of atmospheric evolutionary tracks is shown in Figure 10 where the initial $H_2$ partial pressure is taken at 300 bar, which corresponds to the delivery of $H_2$ associated with the present ocean mass of $H_2O$ under equilibrium with accreting metallic Fe and mantle FeO (Kuramoto and Matsui, 1996). While the atmosphere is $H_2$-dominated, CO and $C_2H_6$ are produced from $CH_4$ slowly due to effective $CH_4$ reformation. After $CH_4$ becomes dominant as $H_2$ is lost by escape, $CH_4$ and $C_2H_6$ become efficiently photolyzed to generate heavier hydrocarbons. It takes ~300 Myr until the partial pressures of $H_2$ and light hydrocarbons $CH_4$ and $C_2H_6$ decrease below 0.1 bar through their loss by escape, oxidation, and heavier hydrocarbon production. Although the depositions of $H_2CO$, HCN, and $NH_3$ little affect the change in atmospheric mass, their net mass reaches ~$10^{17}$ kg.

Production of heavier hydrocarbon is a major branch of the transformation of C during the evolution of the reduced atmosphere.

The major atmospheric components that remain when $H_2$ and light hydrocarbons become largely lost are heavier hydrocarbons regardless of the initial $CH_4/H_2$ ratio (Figure 11(a)) since $CH_4$ tends to be converted to hydrocarbons by their UV shielding effect. The final partial pressures of CO and $CO_2$ are a little large when the initial $CH_4/H_2$ ratio is small because CO and $CO_2$ are major photochemical products in the long-lasting $H_2$-dominated phase (Figure 4(a); Figure 10).

When the initial $CH_4/H_2$ ratio $>\sim 1$, the duration time of the atmospheric state enriched in $H_2$ and/or light hydrocarbons (Figure 11(c)) is almost constant ~10 Myr, which is determined by the time for the decomposition of $CH_4$ and $C_2H_6$. The duration time increases as the initial atmosphere is $H_2$-rich due to the prolongation of time for $H_2$ loss. The formation of a massive $H_2$-rich atmosphere with a mass of several hundred bars on proto-Earth has been suggested from the chemistry of Earth's building blocks involving metallic Fe (e.g., Dauphas, 2017). Late accretion may have produced $H_2$ with a mass comparable to that of the seawater through the accretion of enstatite-chondrite-like materials with a total mass of ~0.5 % of Earth's mass (e.g., Genda et al., 2017; Zahnle et al., 2020; Itcovitz et al., 2022).

The continuous deposition of $H_2CO$ and HCN during the long-lasting atmospheric evolution results in their total supply of $\sim 10^{17}$ kg and $\sim 10^{14}$ kg, respectively (Figure 11(b)). This is equivalent to an $H_2CO$ concentration of ~1 M and an HCN concentration of ~0.01 M in the current volume of seawater; the former is much larger than the minimum $H_2CO$ concentration $\sim 10^{-2}$-$10^{-3}$ M required for the occurrence of formose-type reactions (e.g., Bada and Miller, 1968; Reid and Orgel, 1967; Gabel and Ponnamperuma, 1967; Schwartz and De Graaf, 1993), and the latter is near the level at which HCN hydrolysis and oligomerization may be competing in prebiotic ocean (e.g., Sanchez et al., 1966; Miyakawa et al., 2002). The total deposition of $NH_3$ is small, but $NH_3$ can be produced through the hydrolysis of HCN (Zahnle, 1986; Tian et al., 2011) and the chemical reduction of $N_2$ by accreting metallic Fe (Zahnle et al., 2020; Wogan et al., 2023). $HCONH_2$ can also be produced through the hydrolysis of HCN (e.g., Saladino et al., 2012). The continuous supply of these prebiotically important molecules could potentially lead to the synthesis of amino acids, nucleobases, sugars, and their polymers. When combined with the hot spring hypothesis (Damer and Deamer, 2020), our results indicate that high concentrations of these organic compounds in hot spring pools can be achieved through atmospheric synthesis, as well as other supply processes such as organics production in hydrothermal vents (e.g., Martin et al., 2008) and exogenous delivery (e.g., Chyba and Sagan, 1992). Although part of the deposited organic compounds suffered from several decomposition processes such as hydrolysis and photolysis (Cleaves, 2008; Catling and Kasting, 2017; Pearce et al., 2022), these results suggest that the early atmosphere served as a major source of the prebiotically important molecules.

The efficient production of heavier organics with a total mass equivalent to ~10 bar (Figure 11(a)) may also have played essential roles in the supply of prebiotic compounds. In $CH_4$-dominated conditions, ~80% of organics with 3 carbons are expected to be polymerized to organic haze aerosols supposing that the branching ratio of their production from $C_3H_m$ can be approximated by $R_{pol}/(R_{pol} + R_{ox} + R_{ph})$, where $R_{ph}$ represents the column-integrated photolysis rate of $C_3H_8$, which is the major species among $C_3H_m$, and $R_{pol}$ and $R_{ox}$ represent the column-integrated reaction rate between

CH$_x$ and C$_3$H$_8$, and between oxidant radicals and C$_3$H$_8$ respectively (Supplementary Information for details). Depletion of C$_3$H$_m$ due to their conversion to other species little affects the efficient production of heavier organics by self-shielding because it works even at smaller C$_3$H$_m$ concentrations estimated including the formation of higher hydrocarbons (Supplementary Information). As the water-insoluble organic haze particles gravitationally settle and accumulate on the surface, a global oil slick hundreds of meters thick could have been formed on the surface of the proto-ocean (Lasaga et al., 1971). Such highly polymerized hydrocarbons could supply amino acids and nucleobases through their hydrolysis reactions (Khare et al., 1986; Poch et al., 2012; Pearce et al., 2022). Although nitrogen incorporation into organic haze materials is likely weak due to the strength of the N$_2$ band against photolysis, other energy sources such as lightning, high-energy particles, and impact shocks can decompose N$_2$ to produce N-containing organics (Miller, 1953; Chyba and Sagan, 1992; Furukawa et al., 2009; Ferus et al., 2020; 2022). In addition, the oil slick may have acted as a dry solvent for water-sensitive reactions by preventing the hydrolysis and photolysis of precursor and/or polymerized organics (Lasaga et al., 1971; Nilson, 2002). At the oil-water interface, amino acids and other monomers could become strongly concentrated to produce biopolymers such as peptides (Nilson, 2002).

An atmospheric evolutionary scenario estimated by our calculation is schematically shown in Figure 12. First, an H$_2$-dominated environment lasted for up to several hundred million years. Just after most H$_2$ escaped to space, the production of heavier organics including complex organic haze aerosols became intensified through the enhanced CH$_4$ photolysis and hydrocarbon self-UV shielding. The accumulation of such organics and the prebiotically essential molecules such as HCN and H$_2$CO at this phase can produce various prebiotic compounds as mentioned above. Eventually, the production of biologically essential matters originating from the early atmosphere may have led to the emergence of living organisms.

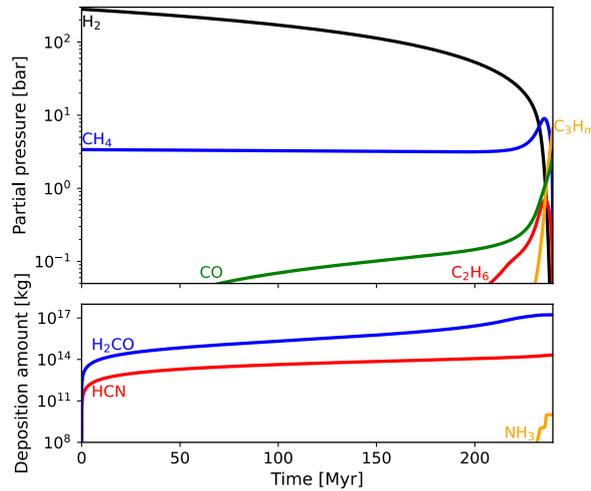

Figure 10. Possible change in the partial pressure and deposition amount of each chemical species with time when the initial $H_2$ partial pressure is 300 bar.

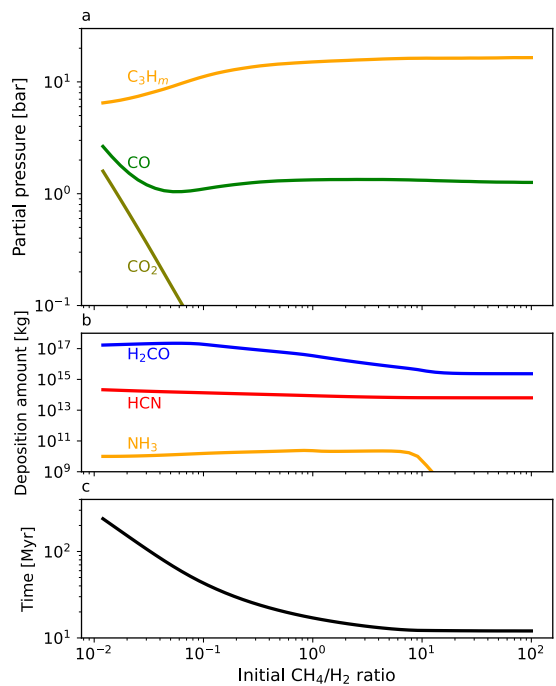

Figure 11. (a) The partial pressure of each chemical species when the partial pressures of $H_2$, $CH_4$, and $C_2H_6$ reach 0.1 bar as a function of the initial $CH_4/H_2$ ratio. (b) The total deposition amounts of HCN, $H_2CO$, and $NH_3$ with the initial $CH_4/H_2$ ratio. (c) Time until the partial pressures of $H_2$, $CH_4$, and $C_2H_6$ reach 0.1 bar as a function of the initial $CH_4/H_2$ ratio.

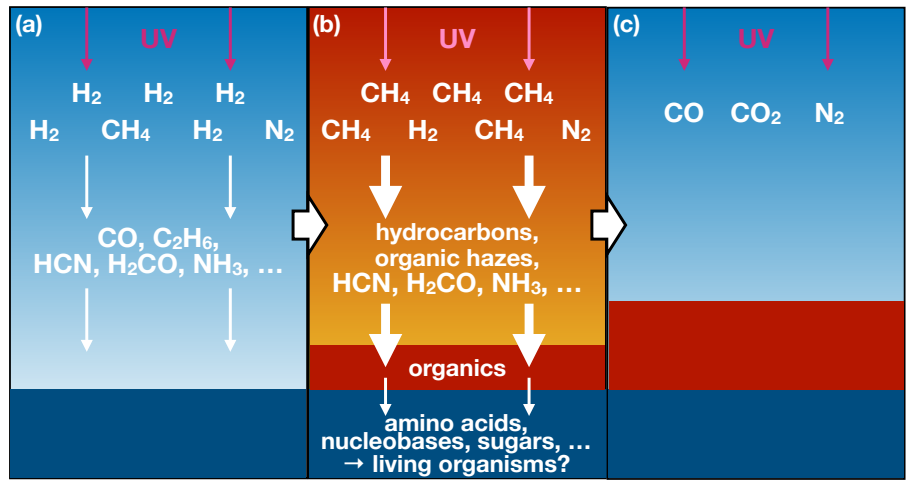

Figure 12. An atmospheric evolution scenario estimated by our calculation. (a) $H_2$-dominated phase where the photochemical production from $CH_4$ is limited. (b) $CH_4$-dominated phase where the production and accumulation of organics are efficient. (c) $CH_4$-decomposed phase where carbon oxides and accumulated organics are left behind.

5. Conclusion

We applied our 1-D photochemical model to Earth's early reduced atmosphere, which was mainly composed of $H_2$ and $CH_4$, to clarify the UV shielding effects of gaseous hydrocarbons on the production of carbon oxides and organic matter. According to our results, UV absorptions by gaseous hydrocarbons such as $C_2H_2$ and $C_3H_4$ significantly suppress the $H_2O$ photolysis and following $CH_4$ oxidation. Accordingly, nearly half of initial $CH_4$ possibly becomes converted to heavier organics along with deposition of prebiotically essential molecules such as HCN and $H_2CO$ on the surface of a primordial ocean for a geological timescale order of 10-100 Myr. These results suggest that the accumulation of organics and prebiotically important molecules in the proto-ocean could have produced a soup enriched in various organics, which might have eventually led to the emergence of living organisms.

**Acknowledgments**

We thank the anonymous reviewer whose comments greatly improved the manuscript. This work was supported by JSPS KAKENHI grant Nos. JP23KJ0093, JP23H04645, JP22KK0044, JP22H00164, JP22KJ0314, JP24KJ0066, JP21K03638, JP24K07110, JP22K21344, and the International Joint Graduate Program in Earth and Environmental Sciences, Tohoku University (GP-EES).

**Figure legends**

Figure 1. Eddy diffusion coefficient profile.

Figure 2. Downward flux of N depending on the basal $CH_4/H_2$ ratio.

Figure 3. Number density profiles when the basal $CH_4/H_2$ ratio is 0.1 (a) and 10 (b).

Figure 4. (a) Column-integrated production/loss rates of $CH_4$, hydrocarbons, and oxides depending on the basal $CH_4/H_2$ ratio. The black dashed line represents the net loss rate of $CH_4$. The solid lines represent the net production rate of each chemical species. "$C_3H_m$" represents the hydrocarbons with 3 carbons. (b) Column-integrated rates of the chemical reactions related to the production (red solid lines) and loss (blue dashed lines) of $CH_4$. (c) Deposition rates of $H_2CO$, HCN, and $NH_3$.

Figure 5. (a) Column-integrated photolysis rate of major UV absorbers. (b) Absorption cross-sections with UV wavelength.

Figure 6. (a) Column-integrated production rates of each major species as a function of the basal $CH_4/H_2$ ratio. The solid lines represent the production rates with UV absorption by hydrocarbons included, while the dashed lines represent the production rates without UV absorption by hydrocarbons other than $CH_4$. (b) Same as (a) but the results without UV absorption by hydrocarbons with 3 carbons ($C_3H_m$). The solid lines are the same as those on (a) and the dashed lines are the results without the UV absorption by $C_3H_m$.

Figure 7. (a) Column-integrated production/loss rates of each chemical species depending on the basal $C_2H_6/CH_4$ ratio. The solid and dashed lines represent the net production and loss rates, respectively. "$C_3H_m$" represents the hydrocarbons with 3 carbons. (b) Deposition rates of HCN, $H_2CO$, and $NH_3$ depending on the basal $C_2H_6/CH_4$ ratio.

Figure 8. (a) Column-integrated production/loss rates of each major species depending on the basal $CO_2/CH_4$ ratio. Here, the basal $CH_4/H_2$ ratio is 10. The solid and dashed lines represent the net production and loss rates, respectively. (b) Same as (a) but the results without UV absorption by hydrocarbons with 3 carbons ($C_3H_m$). The solid lines are the same as those on (a) and the dashed lines are the results without the UV absorption by $C_3H_m$.

Figure 9. (a) Column-integrated production/loss rate when the magnitudes of the eddy diffusion coefficient are the standard setting (Solid lines) and 10 times as large as the standard setting (Dashed lines). (b) Column-integrated production/loss rates when the stratospheric $H_2O$ mixing ratios are 1 ppm (Solid lines) and 10 ppm (Dashed lines). (c) Column-integrated production/loss rates under the standard UV flux setting (Solid lines)

and those under the UV flux half as large as the standard setting (Dashed lines).

Figure 10. Possible change in the partial pressure and deposition amount of each chemical species with time when the initial $H_2$ partial pressure is 300 bar.

Figure 11. (a) The partial pressure of each chemical species when the partial pressures of $H_2$, $CH_4$, and $C_2H_6$ reach 0.1 bar as a function of the initial $CH_4/H_2$ ratio. (b) The total deposition amounts of HCN, $H_2CO$, and $NH_3$ with the initial $CH_4/H_2$ ratio. (c) Time until the partial pressures of $H_2$, $CH_4$, and $C_2H_6$ reach 0.1 bar as a function of the initial $CH_4/H_2$ ratio.

Figure 12. An atmospheric evolution scenario estimated by our calculation. (a) $H_2$-dominated phase where the photochemical production from $CH_4$ is limited. (b) $CH_4$-dominated phase where the production and accumulation of organics are efficient. (c) $CH_4$-decomposed phase where carbon oxides and accumulated organics are left behind.

of impact-generated reduced atmospheres of early Earth. *The Planetary Science Journal*, *1*(1), 11.